\documentclass[aps,preprint,showpacs]{revtex4}

\usepackage{graphicx}
\usepackage{bm}
\begin{document}
\setcounter{page}{1}

\title{Fractional phenomena of the spontaneous emission of a two-level atom in photonic crystals}

\author{Szu-Cheng \surname{Cheng$^{1}$}}
\email {sccheng@faculty.pccu.edu.tw}
\thanks{FAX: +886-2-28610577}
\author{Jing-Nuo \surname{Wu$^{2}$}}
\author{Ming-Rung \surname{Tsai$^{3}$}}
\author{Wen-Feng \surname{Hsieh$^{3}$}}
\email {wfhsieh@mail.nctu.edu.tw}
\affiliation{$^{1}$Department of Physics, Chinese Culture University, Taipei, Taiwan, R. O. C.}
\affiliation{$^{2}$Department of Applied Physics, National Chiayi University, Chiayi, Taiwan, R. O. C.}
\affiliation{$^{3}$Department of Photonics and Institute of Electro-Optical Engineering, National Chiao Tung University, Hsinchu, Taiwan, R. O. C.}
\date[]{Received \today }

\begin{abstract}
We suggest a better mathematical method, fractional calculus, for studying the behavior of the atom-field interaction in photonic crystals. By studying the spontaneous emission of an atom in a photonic crystal with one-band isotropic model, we found that the long-time inducing memory of the spontaneous emission is a fractional phenomenon. This behavior could be well described by the fractional calculus. And the results show no steady photon-atom bound state for the atomic resonant transition frequency lying in the proximity of allowed band edge which is encountered in the previous study [J. Opt. B: Quantum Semiclass. Opt. {\bf 5}, R43 (2003)]. The correctness of this result is validated by the ``cut-off smoothing'' density of photon states (DOS) with fractional calculus. By obtaining a rigorous solution without the multiple-valued problem for the system, we show the method of fractional calculus has logically concise property.
\end{abstract}

\maketitle

\section{INTRODUCTION}
     The performance of photonic devices in various fields is greatly limited by spontaneous emission rate. In light-emitting diodes and lasers, for example, spontaneous emission that is not extracted from the devices will contribute to loss and noise. Inhibiting undesirable spontaneous light emission and redistributing the energy into useful forms becomes important in these fields \cite{22,23,24,25,26}. It has been demonstrated theoretically and experimentally \cite{27, 28} that photonic bandgap (PBG) materials could be effectively used to inhibit the spontaneous emission. Near a photonic band edge, the photon density of state (DOS), which determines the rate of spontaneous emission, is significantly different from that of free space. Singularity \cite{5} of the DOS near PBG leads to the strong atom-field interaction and formation of photon-atom bound states \cite{4a,4b}, where the spontaneous emission rate is inhibited. The Markov approximation \cite{11} of spontaneous emission in free space is no longer valid in this near PBG region, where the atomic decay becomes non-exponential and the emission spectrum becomes non-Lorentzian. By using two-dimensional (2D) photonic crystals, Fujita et al. \cite{29} successfully inhibited and redistributed the spontaneous light emission by a factor of 5 as a result of the 2D photonic bandgap effect.\newline
\indent The experimental data clearly indicate that the spontaneous emission rate has non-exponential decaying behavior when the emission peak is located near the band edges (lattice constant a=480 and 390 nm in Fig. 3) of Ref. \onlinecite{11}. This non-Markov behavior of the PBG reservoir had been  studied by John et al. \cite{4a,4b,9} using the Laplace transform method to solve the time evolution integral equation of the excited probability amplitude of an atom in a high-Q microcavity with singular DOS.\newline
\indent They showed that the time evolution of the excited-state population exhibits decay and oscillatory behavior before reaching a nonzero steady-state value due to photon localization \cite{4a,4b}.  This bound dressed state leads to the fractionalized steady-state atomic population in the excited state. This behavior is observed as the prolonged lifetime effect in Ref. \onlinecite{29}. However, John et al. \cite{9,10} predicted that the unphysical bound state is present even when the resonant atomic frequency lies outside the band gap. This is inconsistent with the experimental result that the prolonged-lifetime effect will disappear when emission peak lies outside the PBG region. This inconsistency may be caused by the multiple-valued problem countered in the studies \cite{9} and by the singular DOS which is appropriate for high-Q microcavity but not suitable for PBG reservoir with more smooth DOS near band edge.\newline  
\indent The time evolution of the probability amplitude of excited level of an atom is related to the delay Green function or memory kernel $G(t-t')$ \cite{5, 11}, which is a measure of the reservoir memory on the excited atom.  The resultant Green function depends very strongly on the photon density of states of the relevant photon reservoir.  The density of states near the band edge in the isotropic one-band model has the form of $\rho(\omega)\propto(\omega-\omega_{c})^{-1/2}$, where the square-root singularity is a characteristic of a one-dimensional phase space.  The corresponding memory kernel \cite{5} has the same form of square-root singularity: $G(t-t')\propto(t-t')^{-1/2}$.  Such a square-root singularity makes the resultant memory kernel possessing the long-time memory effect.   That is, there is no time scale to separate the microscopic levels from the macroscopic levels. The long-time memory phenomena have attracted a great attention in statistical physics. This stochastic property of temporal degree of freedom can be well described by fractional calculus \cite{13a,13b, 14}. The fractional calculus becomes popular in which a diffusion process generated by a fluctuation with no time scale at the macroscopic level. This anomalous diffusion process can be further described by fractional Langevin equation \cite{15, 16}.  \newline
\indent In this paper, we applied the fractional calculus to study the dynamics of the spontaneous emission of an atom in a photonic crystal.  We derived a fractional Langevin equation for this system and solved it to obtain the excited-state probability density. The solution produced by he fractional (inverse) Laplace transform \cite{18a,18b,18c} is expressed in terms of the square complex variables. There is no multiple-valued problem encountered in the previous studies \cite{9,10}. This rigorous mathematical method shows that no steady photon-atom bound state exists for the atomic resonant transition frequency lying in the allowed band. We verified the correctness of this fact by using the ``cut-off smoothing'' DOS \cite{17} with fractional calculus for the atomic transition frequency lying in the proximity of the allowed band edge. The excited-state probability of this result still show decaying characteristics. Fractional calculus gives the correct description of the behavior for the system near the band edge either with or without the ``cut-off smoothing'' DOS. It not only resolves the multiple-valued problem but also avoids choice of smoothing parameter. Therefore, we suggest that the behavior of the atom-field interaction in photonic crystals should be expressed in terms of the fractional calculus.

\section{THE DYNAMICS OF THE SPONTANEOUS EMISSION}

The system we investigate is a two-level atom coupled to the radiation in a photonic crystal with one-band isotropic model. In a rotating-wave approximation, the total Hamiltonian for the coupled atom-field system can be written as

\begin{equation}	H=\hbar\omega_{21}\sigma_{22}+\sum_{\vec{k}}{\hbar\omega_{\vec{k}}a_{\vec{k}}^{+}a_{\vec{k}}+i\hbar\sum{g_{\vec{k}}(a_{\vec{k}}^{+}\sigma_{12}-\sigma_{21}a_{\vec{k}})}}
\end{equation}
   
where $\sigma_{ij}=|{i}\rangle\langle {j}|$ (\textit{i},\textit{j}=1,2) are the atomic operators for a two-level atom with excited state $|{2}\rangle$, ground state $|1\rangle$, and resonant transition frequency $\omega_{21}$; $a_{\vec{k}}$ and $a_{\vec{k}}^{+}$ are the annihilation and creation operators of the radiation field; $\omega_{\vec{k}}$ is the radiation frequency of mode $\vec{k}$ in the reservoir, and the atom-field coupling constant $g_{\vec{k}}=\frac{\omega_{21}d_{21}}{\hbar}[{\frac{\hbar}{2\epsilon_{0}\omega_{\vec{k}}V}]}^{\frac{1}{2}}\hat{e}_{\vec{k}}\cdot\hat{u}_d$ is assumed to be atomic position-independent with the fixed atomic dipole moment $\vec{d}_{21}=d_{21}\hat{u}_d$, the sample volume $V$, polarization unit vector $\hat{e}_{\vec{k}}$ of reservoir mode $\vec{k}$, and the Coulomb constant $\epsilon_{0}$.

In the single photon sector, the wave function of the system has the form
\begin{equation}
	|\psi(t)\rangle=A(t)e^{-i\omega_{21}t}|2,\{0\}\rangle+\sum_{\vec{k}}{B_{\vec{k}}(t)e^{-i\omega_{\vec{k}}t}|1,\{1_{\vec{k}}\}\rangle}
\end{equation}
with initial condition $A(0)=1$ and $B_{\vec{k}}(0)=0$.  Here $A(t)$ labels the probability amplitude for the atom in its excited state $|2\rangle$ with an electromagnetic vacuum state and $B_{\vec{k}}(t)$ for the atom in its ground state $|1\rangle$ with a single photon in mode $\vec{k}$ with frequency $\omega_{\vec{k}}$.

We got the equations of motion for the amplitudes by projecting the time-dependent Schr$\ddot{o}$dinger equation on the one-photon sector of the Hilbert space as

\begin{equation}
	\frac{d}{dt}A(t)=-\sum{g_{\vec{k}}B_{\vec{k}}(t)e^{-i\Omega_{\vec{k}}t}}
\end{equation}

\begin{equation}
	\frac{d}{dt}B_{\vec{k}}(t)=g_{\vec{k}}A(t)e^{i\Omega_{\vec{k}}t}
\end{equation}
with detuning frequency $\Omega_{\vec{k}}=\omega_{\vec{k}}-\omega_{21}$. By substituting the time integration of Eq. (4) into Eq. (3), we have the time evolving equation of the excited-state probability amplitude

\begin{equation}
	\frac{d}{dt}A(t)=-\int_{0}^{t}{G(t-\tau)A(\tau)d\tau}
\end{equation}
with the memory kernel 
$G(t-\tau)=\sum_{k}{g_{\vec{k}}^{2}e^{-i\Omega_{\vec{k}}(t-\tau)}}=\beta^{\frac{3}{2}}\int{\rho(\omega)e^{-i(\omega-\omega_c)}}d\omega$. We could observe from this equation that the memory kernel is a measure of the reservoir's memory in its previous state. The system evolves proceeding according to this equation.                          

Our studying aim, the dynamics of the spontaneous emission, could be obtained by solving the time evolving equation (5). Here we applied fractional calculus and fractional Laplace transform to solve this equation, which were shown to have the form of fractional Langevin equation. We adopted the isotropic one-band model to determine the memory kernel in equation (5), which is given as $G(t-\tau)=\frac{\beta^{3/2}_{1}}{(t-\tau)^{1/2}}e^{-i[\pi/4-\Delta_{c}(t-\tau)]}$  with $t>\tau$ in the long time limit \cite{5}. Here $\Delta_{c}=\omega_{21}-\omega_{c}$ is the detuning frequency of the atomic resonance frequency $\omega_{21}$ from the band edge $\omega_{c}$ and $\beta^{3/2}_{1}=(\omega^{7/2}_{21}d^{2}_{21})/(12\pi^{3/2}\hbar\epsilon_{0}c^{3})$ is the coupling constant. By substituting this memory kernel to Eq. (5), making a transformation $A(t)=e^{i\Delta_{c}t}C(t)$, and applying the definition of the Riemann-Liouville fractional differentiation operator \cite{18a,18b,18c}, we got the fractional form of the time evolving equation as 
\begin{equation}
\frac{d^{3/2}}{dt^{3/2}}C(t)+i\Delta_{c}\frac{d^{1/2}}{dt^{1/2}}C(t)+\sqrt{\pi}\beta^{3/2}_{1}e^{-i\pi/4}C(t)=-\frac{1}{2\sqrt{\pi}}t^{-3/2},
\end{equation}                            
which is defined to be a \textsl{fractional Langevin equation} of this atom-field interaction system. 

We proceeded to solve the probability amplitude $C(t)$ by performing (inverse) Laplace transform on this fractional Langevin equation. It gave
\begin{equation}
C(t)=\sum^{3}_{n=1}a_{n}\left[E_{t}(-\frac{1}{2},X^{2}_{n})+X_{n}e^{X^{2}_{n}t}\right],
\end{equation} 
where $E_{t}(\alpha,a)=t^{\alpha}\sum^{\infty}_{n=0}\frac{(at)^{n}}{\Gamma(\alpha+n+1)}$ is the fractional exponential function of order $\alpha$, $X_{n}$ are the roots of $X^{3}+i\Delta_{c}X-(i\beta)^{3/2}=0$, and $a_{n}$ are $X_{n}$-related coefficients. $X_{n}$ and $a_{n}$ are expressed as 
\begin{equation}
X_{1}=\beta^{1/2}(\eta_{+}+\eta_{-})e^{i\pi/4},
\end{equation}
\begin{equation}
X_{2}=\beta^{1/2}(\eta_{+}e^{-i\pi/6}-\eta_{-}e^{i\pi/6})e^{-i\pi/4},
\end{equation}
\begin{equation}
X_{3}=\beta^{1/2}(\eta_{+}e^{i\pi/6}-\eta_{-}e^{-i\pi/6})e^{i3\pi/4},
\end{equation}
with  
\begin{equation}
\beta^{3/2}=\beta^{3/2}_{1}\sqrt{\pi},\eta_{\pm}=\left[\frac{1}{2}\pm\frac{1}{2}\sqrt{\left(1+\frac{4}{27}\frac{\Delta^{3}_{c}}{\beta^{3}}\right)}^{1/3}\right]
\end{equation}
and  
\begin{equation}
a_{n}=\frac{X_{n}}{(X_{n}-X_{j})(X_{n}-X_{m})}    (n\neq j\neq m;  n,j,m=1,2,3).
\end{equation}
We have the dynamics of spontaneous emission for the system by plotting the excited-state probability density $P(t)=\left|A(t)\right|^{2}=\left|C(t)\right|^{2}$, which has no multiple-valued problem. This probability amplitude could be further expressed by error function as
\begin{equation}
A(t)=e^{i\Delta_{c}t}\sum^{3}_{n=1}a_{n}(X_{n}+Y_{n}(Erf(\sqrt{X^{2}_{n}t})))e^{X^{2}_{n}t}
\end{equation}
with $Y_{n}=\sqrt{X^{2}_{n}} (n = 1,2,3)$, which is the result of previous studies \cite{9, 10}.  However, this expression has multiple-valued problem because the square roots of complex numbers $X^{2}_{n}$ will introduce multiple-valued complex numbers to $Y_{n}$. When the numerical results are shown, every complex sheet of $X^{2}_{n}$ has to be checked one by one to avoid errors. Obviously, our results of fractional calculus are mathematically rigorous and concise. Besides, our result in Fig. 1 shows that there is no steady photon-atom bound state for the atomic transition frequency lying in the proximity of allowed band edge which is very different from the results of previous studies \cite{9, 10}.

    In order to fortify the accuracy of the result of no unphysical photon-atom bound state in the allowed band, we proceeded through applying the "cut-off smoothing" density of state (DOS) to investigating the behavior of the system near the allowed band edge \cite{17}. As mentioned before, DOS in isotropic single band model $\rho(\omega)\propto(\omega-\omega_{c})^{-1/2}\theta(\omega-\omega_{c})$ with Heaviside step function $\theta$ has a weak singularity for $\omega\rightarrow\omega_{c}$ (near band edge). This singular behavior is treated by "cut-off smoothing DOS" $\rho^{s}(\omega)\propto lim_{\epsilon\rightarrow0}\frac{(\omega-\omega_{c})^{1/2}}{(\omega-\omega_{c}+\epsilon)}\theta(\omega-\omega_{c})$ in realistic photonic crystals \cite{17}. Here $\epsilon$ is smoothing parameter and the superscript \textsl{s} denotes the "cut-off smoothing DOS" case. The excited-state probability amplitude $A^{s}(t)=e^{i\Delta_{c}t}C^{s}(t)$ could be exactly solved by performing Laplace transform on the memory kernel and time evolving equation (5). These procedures gave the Laplace transform of probability amplitude $C^{s}(t)$ as
\begin{equation}
\tilde{C}^{s}(s)=\frac{1}{s+i\Delta_{c}+\tilde{G}(s)}
\end{equation}
with $\tilde{G}(s)=\frac{\beta^{3/2}e^{-i\pi/4}}{\sqrt{s}+\sqrt{i\epsilon}}$. This expression could be further rewritten in terms of the roots, $ X^{s}_{n}(n=1,2,3)$, of $z^{3}+\sqrt{i\epsilon}z^{2}+i\Delta_{c}z+(i\sqrt{i\epsilon}\Delta_{c}-\beta^{3/2}i^{3/4})=0$  as 
\begin{equation}
\tilde{C}^{s}(s)=\frac{\sqrt{s}+\sqrt{i\epsilon}}{\prod^{3}_{n=1}(\sqrt{s}-X^{s}_{n})}=\sum^{3}_{n=1}\frac{a^{s}_{n}}{\sqrt{s}-X^{s}_{n}},
\end{equation}
where             
\begin{equation}
X^{s}_{1}=\beta^{1/2}(\eta^{s}_{+}+\eta^{s}_{-})e^{i\pi/4}-\frac{\sqrt{i\epsilon}}{3}
\end{equation}
\begin{equation}
X^{s}_{2}=\beta^{1/2}(\eta^{s}_{+}e^{-i\pi/6}-\eta^{s}_{-}e^{i\pi/6})e^{-i\pi/4}-\frac{\sqrt{i\epsilon}}{3}
\end{equation}
\begin{equation}
X^{s}_{3}=\beta^{1/2}(\eta^{s}_{+}e^{i\pi/6}-\eta^{s}_{-}e^{-i\pi/6})e^{i3\pi/4}-\frac{\sqrt{i\epsilon}}{3}
\end{equation}
with  
\begin{equation}
\eta^{s}_{\pm}=\left(\frac{\chi}{2}\pm\frac{\sqrt{\xi}}{2}\right)^{1/3}, \chi=1-\frac{2}{3}\frac{\Delta_{c}\sqrt{\epsilon}}{\beta^{3/2}}-\frac{2}{27}\frac{\epsilon^{3/2}}{\beta^{3/2}}, \xi=\chi^{2}+\frac{4}{27}\left(\frac{\Delta_{c}-\epsilon/3}{\beta}\right)^{3}
\end{equation}
and 
\begin{equation}
a^{s}_{n}=\frac{X^{s}_{n}+\sqrt{i\epsilon}}{\left(X^{s}_{n}-X^{s}_{j}\right)\left(X^{s}_{n}-X^{s}_{m}\right)}(n\neq j \neq m;n,j,m=1,2,3)
\end{equation} 
Here again we used fractional calculus (fractional inverse Laplace transform) to obtain the excited-state probability amplitude
\begin{equation}
A^{s}(t)=e^{i\Delta_{c}t}\sum^{3}_{n=1}a^{s}_{n}\left[E_{t}(-\frac{1}{2},(X^{s}_{n})^{2})+X^{s}_{n}e^{(X^{s}_{n})^{2}t}\right]
\end{equation}

As we study how the system behaves near the allowed band edge, we choose the detuning frequency $\Delta_{c}=0.3\beta$ with smoothing parameter $\epsilon=10^{-3},10^{-5},0$, respectively in Fig. 2. It could be observed that the excited-state probability density $P^{s}(t)=\left|A^{s}(t)\right|^{2}$ has small oscillatory behavior in the short time regime but approaches zero in the long time limit. It means that there is really no steady photon-atom bound state for the atomic transition frequency lying in the proximity of allowed band edge. Actually, we have plotted all the behavior of the system inside the allowed band in Fig. 3. For the atomic resonant transition frequency located deep inside the allowed band ($\Delta_{c}/\beta_{1}=10$) or very close to the band edge ($\Delta_{c}/\beta_{1}=0.01$), we found that these probabilities all show decaying characteristics in the long time limit. That is, the photon located within the allowed band will not strongly interact with atom so the photon-atom bound state will not be formed. This result could be verified analytically from the mathematical expression of the excited-state probability amplitude $A^{s}(t)=e^{i\Delta_{c}t}\sum^{3}_{n=1}a^{s}_{n}\left[E_{t}(-\frac{1}{2},(X^{s}_{n})^{2})+X^{s}_{n}e^{(X^{s}_{n})^{2}t}\right]$. For the positive detuning (inside the allowed band $\Delta_{c}>0$), both terms in the square bracket will asymptotically cancel out each other as time approaches infinity ($t\rightarrow\infty$). We get correct depiction of the dynamics of spontaneous emission in a PBG reservoir through fractional calculus, which is proposed as a better mathematical method to study the behavior of the atom-field interaction in photonic crystals.  
    
\section{CONCLUSION}

The dynamics of the spontaneous emission of an atom in a photonic crystal with one-band isotropic band structure can be treated by the fractional calculus using either singular or "cut-off smoothing" density of photon states.  For the first time to our knowledge we show that it is a fractal phenomenon that induces the long-time memory of the spontaneous emission in the photonic crystal.  Solving the time evolving equation of the probability amplitude for the system governed by the fractional memory kernel described by singular density of states, we obtained rigorous solutions without multiple-valued problem encountered.  Besides, we found that there is no unphysical state of fractionalized atomic population in the excited state when the resonant atomic frequency lies in the allowed band, even extremely close to the band edge. This result was validated by the "cut-off smoothing DOS" with fractional calculus. We suggest that the correct description of the dynamics of the spontaneous emission in a photonic crystal should be expressed in terms of the fractional calculus. This mathematical method, concerned about the isotropic model here, can be easily extended to the anisotropic ones and to study interesting effects such as the enhancement of the index of refraction with greatly reduced absorption, electromagnetically induced transparency, and optical amplification without population inversion. 

\section{Acknowledgement}
We would like to gratefully acknowledge partially financial support from the National Science Council (NSC) in Taiwan under Contract Nos. NSC-96-2914-I-009-017, NSC-96-2628-M-009-001, and NSC-95-2119-M-009-029.

\bibliography{Fractional}

\begin{figure}[t]
\includegraphics[width=12.0 cm]{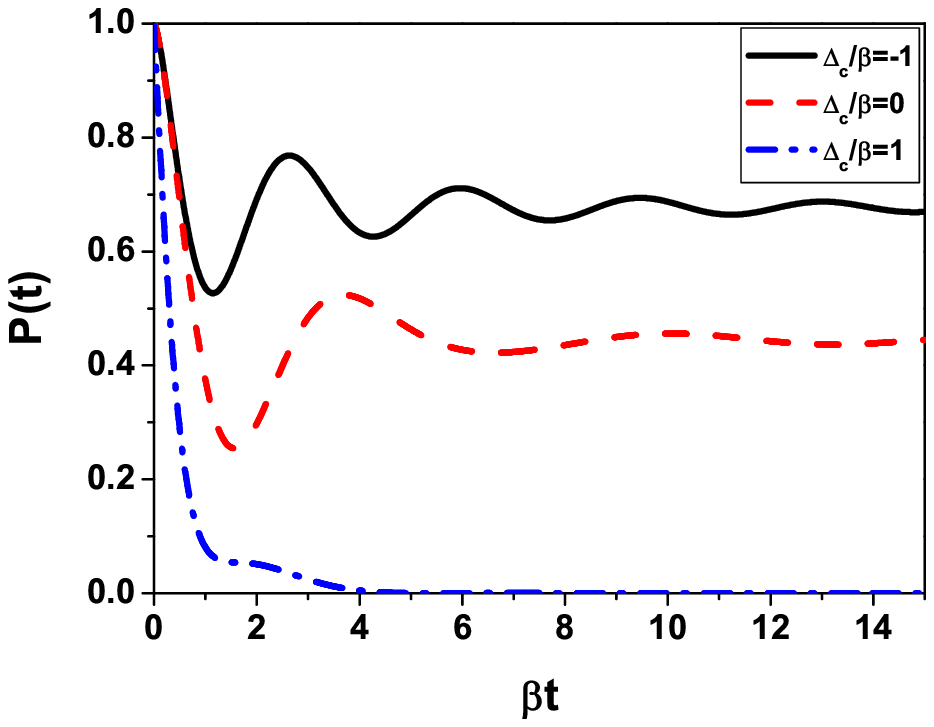}\label{fig 1}
\caption{Excited-state probability density, $P(t)=|A(t)|^{2}$, as a function of $\beta t$ for various values of the atomic detuning frequency ($\Delta_{c}=\omega_{21}-\omega_{c}$), $\Delta_{c}/\beta=-1$ (inside the band gap), $\Delta_{c}/\beta=0$ (at the band edge) and $\Delta_{c}/\beta=+1$ (within the allowed band)}
\end{figure}

\begin{figure}[b]
\includegraphics[width=12.0 cm]{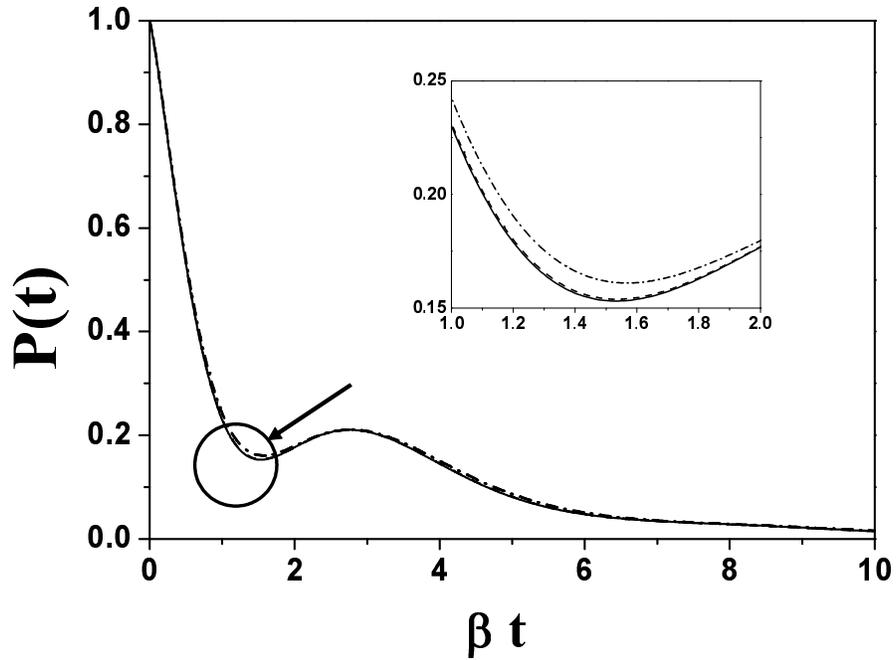}\label{fig 2}
\caption{Excited-state probability density, $P^{s}(t)=\left|A^{s}(t)\right|^{2}$, for the atomic detuning frequency $\Delta_{c}/\beta=0.3$ with three values of smoothing parameter $\epsilon=0$ (solid line), $\epsilon=10^{-5}$ (dashed line),$\epsilon=10^{-3}$ (dot dashed line). The difference of these lines marked by circle is enlarged and shown in the inset}
\end{figure}
\begin{figure}[t]
\includegraphics[width=12.0 cm]{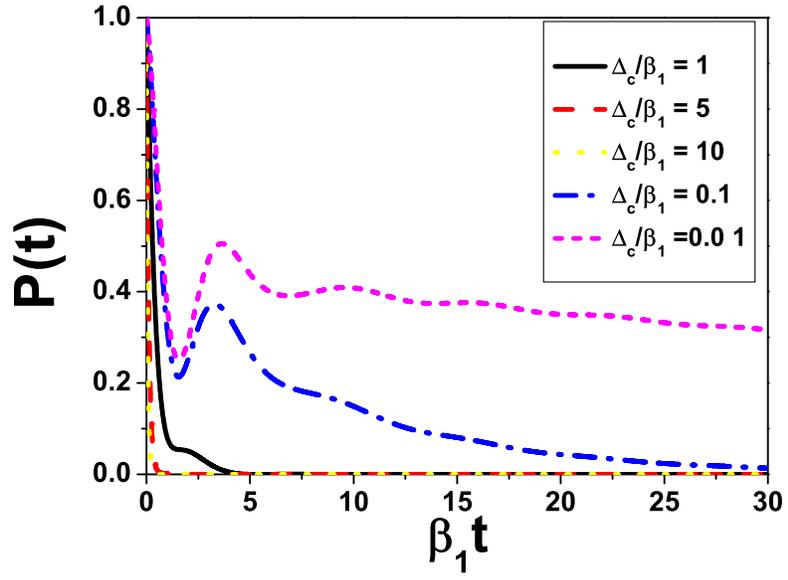}\label{fig 3}
\caption{Excited-state probability density, $P(t)=\left|A(t)\right|^{2}$, for various values of atomic detuning frequency inside the allowed band ($\Delta_{c}/\beta>0$)}
\end{figure}

\end{document}